\documentclass[12pt]{article}

\usepackage{graphicx}

\def \lsim{\mathrel{\vcenter
     {\hbox{$<$}\nointerlineskip\hbox{$\sim$}}}}

\newcommand{\beq}{\begin{equation}}
\newcommand{\eeq}{\end{equation}}
\newcommand{\beqa}{\begin{eqnarray}}
\newcommand{\eeqa}{\end{eqnarray}}
\newcommand{\beqar}{\begin{eqnarray*}}
\newcommand{\eeqar}{\end{eqnarray*}}

\begin{document}
\thispagestyle{empty}

\hfill{\sc UG-FT-282/11}

\vspace*{-2mm}
\hfill{\sc CAFPE-152/11}

\vspace{32pt}
\begin{center}
{\textbf{\Large Heavy-neutrino decays at}}

{\textbf{\Large neutrino telescopes}}

\vspace{40pt}

Manuel Masip, Pere Masjuan
\vspace{12pt}

\textit{
CAFPE and Departamento de F{\'\i}sica Te\'orica y del Cosmos}\\ 
\textit{Universidad de Granada, E-18071, Granada, Spain}\\
\vspace{16pt}
\texttt{masip@ugr.es, masjuan@ugr.es}
\end{center}

\vspace{40pt}

\date{\today}

\begin{abstract}

It has been recently proposed that a sterile neutrino $\nu_h$ 
of mass $m_h=40$--$80$ MeV, mixing 
$|U_{\mu h}|^2\approx 10^{-3}$--$10^{-2}$, lifetime 
$\tau_h\lsim 10^{-9}$ s, and a dominant decay mode 
$\nu_h\rightarrow \nu \gamma$ could be the origin
of the experimental anomalies observed at LSND, KARMEN and
MiniBooNE. Such a particle would be abundant inside air showers, 
as it can be produced in kaon decays 
($K\rightarrow \nu_h\mu$, 
$K_L\rightarrow \nu_h \pi \mu$). 
We use the $Z$-moment method 
to evaluate its atmospheric flux and the frequency of its decays
inside neutrino telescopes. We show that $\nu_h$ would imply  
around $10^4$ contained showers 
per year inside a 0.03 km$^3$ telescope like 
ANTARES or the DeepCore in IceCube.
These events would have a characteristic energy 
and zenith-angle distribution ($E_\nu\approx 0.1$--$10$ TeV 
and $\theta < 90^o$), which results from a 
balance between the {\it reach} of the heavy neutrino 
(that disfavors low energies) and a sizeable production rate 
and decay probability. The standard 
background from contained neutrino events
($\nu_e N\rightarrow e X$ and neutral-current interactions of 
high inelasticity) is $100$ times smaller.
Therefore, although it may be challenging
from an experimental point of view, a search  
at ANTARES and IceCube could 
confirm this heavy-neutrino possibility.


\end{abstract}

\newpage

\section{Introduction}
The direct observation of neutrino interactions 
in different types of experiments \cite{Conrad:2007ea}
has been used to establish a basic picture of neutrino masses 
and mixings. From a model building point of view, this is 
arguably the most significant discovery occurred in
particle physics since the confirmation of the standard model
in the early 70's, as it reveals a scale that is (most likely)
not electroweak. The picture, however, has faced some
persistent anomalies in experiments with neutrino beams from
particle accelerators. Basically, muon
neutrinos of energy below 1 GeV seem to
experience an excess of charged-current (CC) interactions with
an electron in the final state. The interpretation of these 
events in terms
of $\nu_\mu\to \nu_e$ oscillations is inconsistent
with the mass parameters deduced from solar, atmospheric
and reactor neutrino observations.

In a recent analysis Gninenko \cite{Gninenko:2010pr}
has made a very compelling case
for a massive neutrino as the origin for all these anomalies:

{\it (i)} LSND \cite{Athanassopoulos:1996jb}
observed $\bar \nu_e$-like events 
($\bar \nu_e\, p\rightarrow e^+ n$) with 
a gamma signal from neutron capture
that seem to imply an excess of $\bar \nu_\mu\to \bar \nu_e$ 
oscillations. He shows that the events could
be equally explained through $\nu_h$ production 
($\nu_\mu ^{\;12}C\rightarrow \nu_h n X$) 
followed by its radiative 
decay $\nu_h\rightarrow \nu\, \gamma$, with the final $\gamma$ converted
into a $e^+e^-$ pair indistinguishable from an electron.
This explanation would work for a large enough production cross section
($|U_{\mu h}|^2\approx 10^{-3}$--$10^{-2}$ and $m_\nu < 80$ MeV)
and a short enough decay length ($\tau_h\lsim 10^{-9}$ s and 
$m_\nu > 40$ MeV). During the first years of data taking LSND
also observed an excess of $\nu_e ^{\;12}C\rightarrow e^- X$
events that were interpreted as $\nu_\mu\to \nu_e$ oscillations 
but are consistent as well with the $\nu_h$ hypothesis.

{\it (ii)} KARMEN \cite{Armbruster:2002mp}, 
using a similar technique, did not confirm
the LSND anomalies.
The neutrinos at LSND, however, had an average energy of 100 MeV
and a long high-energy tail, whereas the spectrum at KARMEN was
a narrow peak around 20 MeV.
Gninenko shows that a 40 MeV neutrino would be above
the production threshold at KARMEN, which makes his hypothesis 
consistent with the data.

{\it (iii)} MiniBooNE \cite{AguilarArevalo:2007it}
has observed an excess of electron-like
events for $\nu_e$ energies between 200 and 475 MeV, with no significant
excess at higher energies. Gninenko's fit 
exhibits also a good agreement with the data (higher-energy 
events are disfavored by an increase in the decay length
and are hidden by the low statistics). More recently 
\cite{AguilarArevalo:2010wv} this experiment
has also reported an excess in $\bar \nu_\mu$ data for antineutrino
energies in a wider range. His fit is consistent as well, and could favor
a Dirac nature for $\nu_h$.

\noindent 
In addition, the mass range $40\le m_h\le 80 \; {\rm MeV}$ makes 
$\nu_h$ too heavy to be produced in pion decays and too light to distort
the muon spectrum in kaon decays. The heavy neutrino is also produced
when the muon itself decays, but he argues that a mixing 
$|U_{\mu h}|^2 < 10^{-2}$ makes it acceptable. Specific searches for
unstable neutrinos put strong constraints on $|U_{\mu h}|$, but are
based on decays with charged particles in the final state
($\nu_h\rightarrow ee\nu,\mu e \nu,\mu\pi\nu$), never on the decay
$\nu_h\rightarrow \nu\gamma$ induced by a magnetic moment transition.
Its large mass and short lifetime should keep $\nu_h$ also {\it safe} 
from bounds from supernovae and primordial 
nucleosynthesis \cite{Dolgov:2000pj}. 
Finally, a recent analysis \cite{McKeen:2010rx} of 
muon capture with photon emission at TRIUMF finds that  
Gninenko's neutrino would imply a signal well above 
the 30\% excess (versus the standard model value) 
deduced from the data \cite{Bernard:2000et}. 
One should notice, however, that 
the photon energy cut and the small size of the
target volume at TRIUMF make this experiment very sensitive
to the neutrino lifetime. A value $\tau_h\approx 3\times 10^{-9}$ s 
could imply a consistent radiative capture rate there
while explaining the data at LSND and KARMEN (which
require  $\tau_h\le 10^{-8}$ s)
and still having an impact at MiniBooNE\footnote{A global fit 
including TRIUMF would certainly constrain further the parameter 
space in Gninenko's model.}.

We find the heavy neutrino hypothesis very interesting
and will study here its implications in a different type of 
experiments. Our basic observation is
that $\nu_h$ would be abundantly 
produced in the atmosphere through  kaon decays. 
At energies around 1 TeV its decay length
becomes $c\tau_h\gamma \approx 5$ km, which implies that 
$\nu_h$ can reach a neutrino telescope and then decay. The final
photon would be seen there as a {\it pointlike} event, 
similar to the shower from 
$\nu_e N\to e X$ or from a neutral-current (NC) interaction of high
inelasticity but clearly distinguishable from the muon track
in $\nu_\mu N\to \mu X$.

\section{Neutrino fluxes at sea level}
The atmospheric flux of any species can be easily estimated 
using the $Z$-moment method \cite{gaisser90,Lipari:1993hd}. 
This method provides a set of coupled 
differential equations that describe the evolution with the
atmospheric depth $t$ (in g/cm$^2$) 
of the fluxes of {\it parent} hadrons ($\phi_H$ with 
$H=p,n,\pi^\pm,K^\pm,K_L$) and of any particles that may result
from their decay or their collision with an air
nucleus. The generic equations for $\phi_H(E,\theta,t)$ are
\beq
{\partial \phi_H\over \partial t}=-{\phi_H\over \lambda^H_{\rm dec}}
-{\phi_H\over \lambda^H_{\rm int}}+\sum_{H'}S_{H'H}\;,
\label{dfdt}
\eeq
where $\lambda^H_{\rm dec}$ ($\lambda^H_{\rm int}$) is 
the decay (interaction) length of $H$ in the air and $S_{H'H}$
describe the sources. These equations can be solved 
analytically under some simplifying assumptions, namely,\\
{\it (i)} the all-nucleon 
primary flux has a constant spectral index $-\alpha$;\\ 
{\it (ii)} the energy distribution of particles from collisions 
and decays scales linearly with the energy of the parent hadron;\\ 
{\it (iii)} the hadronic interaction lengths $\lambda^H_{\rm int}$
do not change with the energy;\\
{\it (iv)} the contributions to the nucleon flux from 
meson collisions and to the pion flux from kaon collisions
are negligible.\\ 
It follows that the nucleon 
fluxes $\phi_N$ 
keep the same spectral index $-\alpha$ at any depth, and that 
the source terms are reduced to 
\beq
S_{NH}={\phi_N\over \lambda^N_{\rm int}} \, Z_{NH}\;
\label{s}
\eeq
where the $Z$--factors
\beq
Z_{NH}=\int_0^1{\rm d}x\;x^{\alpha-1}F_{NH}\;
\eeq
are constants (independent of $E$ and the zenith angle $\theta$) derived 
from the distribution $F_{NH}(x)$ of the fraction of 
energy taken by $H$ after a $N$--air collision.
The meson fluxes $\phi_M$ can then be easily derived 
in two different regimes. At low energies 
$\lambda^M_{\rm dec}=(E /m_M) c\tau_M \rho$ 
is much smaller than $\lambda^M_{\rm int}$ and meson
interactions can be ignored, whereas at high energies the
variations in $\phi_M$ are dominated by collisions 
with air nuclei (the air density $\rho$ 
is a function of $t$ and of $\theta$).
A simple interpolation can be used between these regimes.
We take in our analysis the primary flux, the $Z$-factors 
and the atmospheric model in \cite{Lipari:1993hd} 
(see \cite{Illana:2010gh} for a discussion 
of the fluxes at higher energies).
We obtain, for example, that the TeV charged-pion vertical 
flux reaches its maximum (a $4\%$ of the initial 
nucleon flux) at a depth of 200 g/cm$^2$, and that the
kaon flux there is 7 times smaller.

The lepton fluxes from meson decays can also be  
incorporated. In particular, standard 
neutrinos do not interact nor decay in the atmosphere 
and their flux equations will only depend on source terms
of type 
\beq
S_{M\nu}(E)=B(M\to \nu)\int_0^1 {\rm d}x \, x^{-1}\,
{\phi_M(E/x)\over \lambda^M_{\rm dec}(E/x)}\,
F_{M\nu}(x)\;,
\label{snu}
\eeq
where $B(M\to \nu)$ is the branching ratio of a given decay 
mode, $F_{M\nu}(x)$ is again the distribution of the fraction of 
energy taken by the neutrino in that decay, and the dependence
on $t$ and $\theta$ is implicit. We obtain that, although kaons
are less abundant than pions in air showers, a
lower ratio $\lambda_{\rm dec}^M/\lambda_{\rm int}^M$ 
makes them the main source of neutrinos at energies above
100 GeV. The TeV flux at sea level is dominated
by muon neutrinos, with 
$\phi_{\bar \nu_\mu}\approx 0.42\, \phi_{\nu_\mu}$, 
$\phi_{\nu_e}\approx 0.036\, \phi_{\nu_\mu}$ and
$\phi_{\bar \nu_e}\approx 0.023\, \phi_{\nu_\mu}$.

\begin{figure}
\begin{center}
\includegraphics[width=0.7\linewidth]{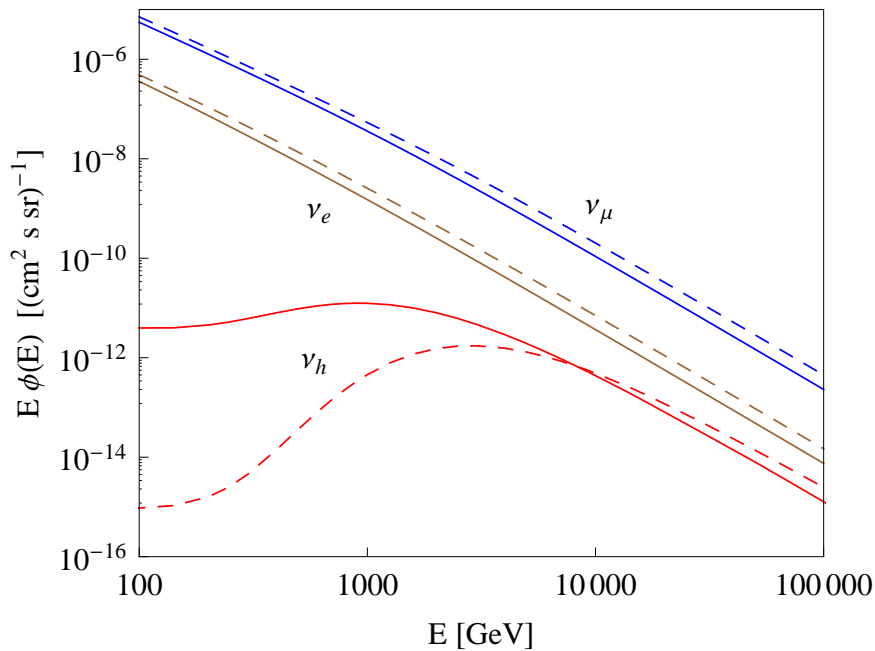} 
\end{center}
\caption{Neutrino fluxes ($\nu_i+\bar \nu_i$) at sea level for
$\theta=0$ (solid) and $\theta=60^o$ (dashes).
\label{fig1}} \end{figure}
Heavy neutrinos $\nu_h$ will be mainly produced 
in charged-kaon decays. The  branching ratio is 
\beq
B(K^+\to \mu^+ \nu_h)\approx B(K^+\to \mu^+ \nu)
\times |U_{\mu h}|^2\;\bar \rho_h\;,
\label{fnh}
\eeq
where $B(K^+\to \mu^+ \nu)=0.64$ and the kinematic factor
for $m_h=40$--80 MeV is $\bar \rho_h\approx (1+m_h^2/m_\mu^2)$ 
\cite{Shrock:1980ct}.
The fraction of energy $x$ taken by $\nu_h$ will have a flat 
distribution (a constant $F_{Kh}(x)$) between $x_{min}$ and
$x_{max}$, 
\beq
x_{\stackrel{min}{\scriptscriptstyle max}}= 
{1\over 2}\left( 1+{m_h^2-m_\mu^2\over m_K^2}\right)
\mp \sqrt{{1\over 4} \left( 1+{m_h^2-m_\mu^2\over m_K^2}\right) -
{m_h^2\over m_K^2}}\;.
\eeq
There will be smaller contributions from $K^+\to \pi^0\mu^+\nu_h$ 
and $K_L\to \pi^-\mu^+\nu_h$, plus the analogous $K^-$ and $K_L$ 
decays into $\bar \nu_h$ (the heavy neutrino may be a Dirac or a
Majorana particle, 
see discussion in \cite{Gninenko:2010pr}).
The equation defining $\phi_h(E,\theta,t)$
is
\beq
{\partial \phi_h\over \partial t}=-{\phi_h\over \lambda^h_{\rm dec}}
+\sum_{K}S_{Kh}\,
\eeq
where the source terms take the form in Eq.~(\ref{snu}), 
$\lambda^h_{\rm dec}=(E /m_h) c\tau_h \rho$, and
the sum runs over the decay modes that produce $\nu_h$.
In Fig.~1 we plot the total heavy neutrino 
flux at sea level from inclinations $\theta=0^o,60^o$. 
We have taken the central values $m_h=60$ MeV and 
$|U_{\mu h}|^2= 0.005$, with 
$\tau_h= 10^{-9}$ s. At 1 TeV 73\% of the flux comes 
from $K^+$ decays, $K^-$ contributes a 25\%, and 
$K_L$ just a 2\%. Finally, notice that the photons produced
in the air through $\nu_h$ decays are together with other 
photons and muons {\it inside} the parent shower
and are therefore non-observable.

\section{Events at a neutrino telescope}
As neutrinos enter the ground their sources disappear and
they just experience two
types of processes: heavy neutrinos $\nu_h$ may decay
into $\gamma\,\nu_\mu$,
whereas $\nu_\mu$ and $\nu_e$ may have neutral-
or charged-current interactions with matter. At a 
depth $d$ the sea-level 
fluxes $\phi_i(E,\theta,0)$ become 
\beqa
\phi_h(E,\theta,d)&=&\phi_h(E,\theta,0) \; 
\exp\left({-d\over \lambda_{\rm dec}^h\cos\theta}\right)\;;
\nonumber \\
\phi_\nu(E,\theta,d)&=&\phi_\nu(E,\theta,0) \;
\exp\left({-d\over \lambda_{\rm int}^\nu\cos\theta}\right)\;,
\eeqa
where $d$ and the decay/interaction lengths are given in
meters and we have neglected the curvature of the Earth
(a good approximation for $\theta\le 85^o$).
At 1 TeV we obtain $\lambda_{\rm dec}^h\approx 5$ km whereas
$\lambda_{\rm CC}^\nu\approx 2\times 10^6$ km and 
$\lambda_{\rm NC}^\nu\approx 8\times 10^6$ km (the interaction
lengths decrease with the energy as $1/\sigma_{\nu N}$). 
This means
that the decay of a $\nu_h$ crossing a neutrino telescope is 
$10^6$ times more probable than the interaction of 
a standard neutrino.
In addition, CC $\nu_\mu$  interactions will
be clearly different from $\nu_h$ decays, as the final muon
will produce a track hundreds of meters long \cite{Halzen:2010yj}. 
The electromagnetic
shower from a $\nu_h$ decay will be pointlike (it develops
in a few meters), similar to the one produced by 
a NC  interaction or a $\nu_e$ CC process. 
These standard events, however, are 
suppressed by the lower $\nu_e$ fluxes ($\nu_\mu\to \nu_e$
oscillations at 
$L\le 100$ km and $E\ge 100$ GeV are negligible) and the
inelasticity distribution ($\propto 1/y$ \cite{Connolly:2011vc}) 
in $\nu$--$N$ collisions.

\begin{figure}
\begin{center}
\includegraphics[width=0.5\linewidth]{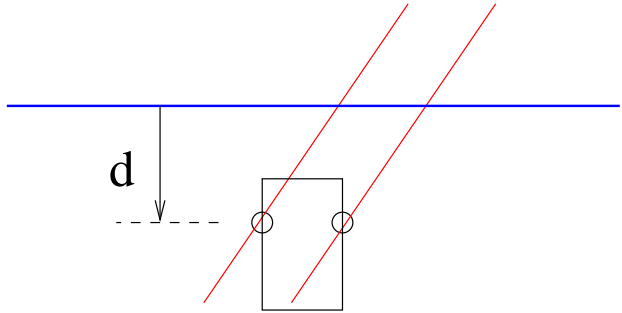} 
\end{center}
\caption{The total flux through the lateral surface of the
detection region cancels if $R_T\gg d/\cos\theta$ (we take
$\theta\le 85^o$).
\label{fig2}}
\end{figure}
Let us be more specific. To estimate the number of events per
unit time 
occurred inside a telescope one needs to calculate the total
flux (ingoing plus outgoing neutrinos) through the surface 
containing the detectors. 
We model this region as a cylinder of section $A$
and length $H$ starting at a depth $d_0$ ({\it i.e.}, $d$ goes
from $d_0$ to $d_0+H$).
For a fixed angle $\theta\lsim 85^o$ the neutrino flux 
only depends on the depth $d$. 
Therefore, the total 
flux through the lateral surface of the detector will be zero (given
$\theta$, the flux through any lateral ${\rm d}\vec S$ is
equal to the flux leaving the detector through an opposed lateral 
surface $-{\rm d}\vec S$, see Fig.~2). The 
number of heavy-neutrino events inside the
detector in an interval of energy and
solid angle per unit time can then be calculated as the difference
between the fluxes through its upper and its lower sections:
\beqa
N_h=\int_{\Delta E} \! {\rm d}E 
\int_{\Delta \Omega}\!  {\rm d}\Omega \int_{A}  {\rm d}S\; & 
\!\! \cos\theta \;
[\, \phi_h\!\left(E,\theta,d_0\right) \nonumber \\
\!- &\! \phi_h\!\left(E,\theta,d_0+H\right)\, ]\,.
\eeqa
An analogous expression can be obtained to estimate the number
$N_\nu$ of interactions inside the detector produced by neutrinos
from those directions.
In Fig.~3 we plot in dashes the energy 
distribution of the neutrinos that interact ($\nu_\mu$ and $\nu_e$)
or decay ($\nu_h$) per year 
inside a detector like ANTARES 
\cite{Brunner:2011zz} ($A=0.1$ km$^2$, 
$d_0=2.2$ km, $H=0.3$ km) or the DeepCore \cite{Wiebusch:2009jf}
in IceCube (of similar size and depth). 
\begin{figure}
\begin{center}
\includegraphics[width=0.7\linewidth]{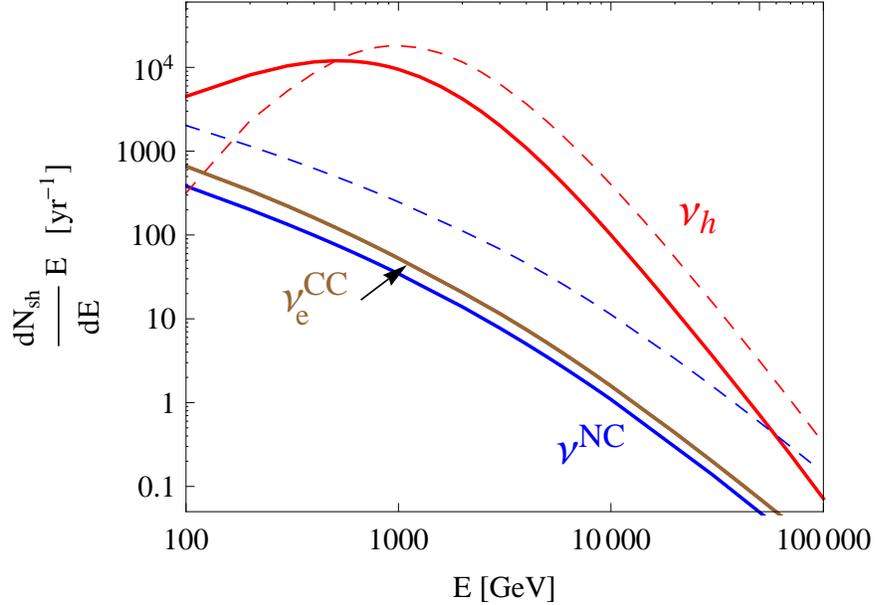} 
\end{center}
\caption{Energy distribution of the contained showers from 
$\nu_h\to \nu_\mu\gamma$, $\nu N\to \nu X$ and 
$\nu_e N\to e X$ at ANTARES. We also plot (dashes) the
energy distribution of the parent neutrino in each case.
\label{fig3}} 
\end{figure}
The energy of the initial neutrino,
however, is not the most relevant parameter for observation,
as in NC interactions only a small fraction $y$ may be
deposited in the detector. In a 
$\nu_h$ decay only the photon energy is {\it visible} (we will
assume an isotropic decay \cite{Gninenko:2010pr}), whereas 
in $\nu_e$-CC interactions all the energy carried by the neutrino
goes to the contained shower. If the inelasticity $y$ 
in the event has a 
distribution $F_{\nu\,  {\rm sh}}(y)$, then the energy distribution
of the cascades inside the detector will be
\beq
{{\rm d}N_{\rm sh}\over {\rm d}E}(E)=
\int_0^1 {\rm d}y\;y^{-1}\,{{\rm d}N_{\nu}\over {\rm d}E}\!(E/y)\,
F_{\nu\, {\rm sh}}\!(y)\;.
\eeq
Fig.~3 shows in solid lines 
the main result of our analysis. The number of
standard showers of energy above 100 GeV from 
down-going neutrinos inside ANTARES is
around 1300 per year (60\% from $\nu_e$-CC interactions 
and 40\% from NC interactions).
For the central values $m_h=60$ MeV and $|U_{\mu h}|^2=0.005$ 
 the heavy neutrino would 
provide 26000 extra events. If the energy
threshold is set at 500 GeV the number of standard events
is reduced to 220 per year, whereas the number of events from 
$\nu_h$ decays is just cut to 14000.

\section{Summary and discussion}
Telescopes like ANTARES or IceCube are designed to observe 
upward-moving muons produced in neutrino interactions 
near the detector. These events are {\it clean}, 
in the sense that no particles except for neutrinos
can reach the detector after crossing the Earth. Telescopes can 
also observe the contained showers produced in 
NC interactions or 
in $\nu_e$-CC processes. Since their development takes just a few
meters, these events are pointlike, 
and the only sign indicating whether they are caused by an upward or a
downward-going neutrino is that in the latter case they may come together 
with muons.

In this paper we have shown that the decay of a long-lived neutral
particle produced in the atmosphere could change drastically (by 
over a factor of 100) the number of TeV contained showers in these
experiments. In particular, we have analyzed 
Gninenko's heavy neutrino, that 
appears as a possibility well
motivated by the results at LSND, KARMEN and MiniBooNE.
We find remarkable that its mass, mixing, and  lifetime  optimize the 
{\it distortion} 
introduced in TeV-neutrino telescopes: below $\approx 100$ GeV 
$\nu_h$ does not reach the telescope, and above 100 TeV its decay length
becomes too large and the signal vanishes 
(notice that $\lambda_{\rm dec}^h$ grows 
with the energy  while $\lambda_{\rm int}^\nu$ decreases).

The heavy neutrino would be produced through kaon decays inside  
air showers together with a muon of similar energy. 
A crucial question is then whether these 
decays can be disentangled from the muon bundle 
associated to the parent shower. ANTARES or the DeepCore
in IceCube are more than 2 km deep, and as the zenith angle
grows all muon effects will decrease. In contrast, the 
zenith angle dependence of the $\nu_h$ events up to $85^o$ 
is very mild (especially at larger energies), since
these neutrinos do not lose energy in their way to the
detector. Therefore, the TeV contained showers should appear as
a clear anomaly that may be accompanied by muons in vertical
events but that
{\it persists} at higher zenith angles (slant depths).
The large number of events that we obtain could allow 
for specific searches. For example, 
events with lower-energy 
muons in the upper part of the detector followed by a contained TeV
shower below, or other topologies that would otherwise be
discarded.

If a heavy neutrino is the explanation of the  
LSND and MiniBooNE anomalies, our results show that it will 
reach effectively the core
of neutrino telescopes and will decay there at a high rate.
A MonteCarlo simulation of individual showers, including all
the muon backgrounds and the response of the detector, 
should then provide the best strategy in the search for
an observable signal that 
confirms or excludes this heavy neutrino possibility.

\section*{Acknowledgments}
We would like to thank Juande Zornoza for discussions.
This work has been partially supported by
MICINN of Spain (FPA2006-05294, FPA2010-16802, FPA2010-16696,  
Consolider-Ingenio 
{\bf Multidark} CSD2009-00064 and 
{\bf CPAN} CSD2007-00042)  
and by Junta de Andaluc\'{\i}a
(FQM 101, FQM 437 and FQM 3048).

\end{document}